# Measurement-Based Estimation of System State Matrix for AC Power Systems with Integrated VSCs

Jinpeng Guo, *Student Member, IEEE*, Xiaozhe Wang, *Member, IEEE*, Boon-Teck Ooi, *Life Fellow, IEEE*

*Abstract*—In this paper, a wide-area measurement system (WAMS)-based method is proposed to estimate the system state matrix for AC system with integrated voltage source converters (VSCs) and identify the electromechanical modes. The proposed method is purely model-free, requiring no knowledge of accurate network topology and system parameters. Numerical studies in the IEEE 68-bus system with integrated VSCs show that the proposed measurement-based method can accurately identify the electromechanical modes and estimate the damping ratios, the mode shapes, and the participation factors. The work may serve as a basis for developing WAMS-based damping control using VSCs in the future.

## NOMENCLATURE

| | |
|---|---|
| $M$ | Generators' inertia coefficient matrix |
| $D$ | Generators' damping coefficient matrix |
| $\delta$ | The vector of generators' rotor angles |
| $\omega$ | The vector of generators' rotor speeds |
| $E$ | A diagonal matrix whose diagonal components are generators' voltage magnitude behind transient reactance |
| $G$ | A diagonal matrix whose diagonal entries are the diagonal elements of the reduced conductance matrix |
| $\Sigma$ | The load fluctuation intensity matrix |
| $\xi$ | A standard Gaussian random vector |
| $A$ | Open-loop dynamic system state matrix |
| $A_c$ | Closed-loop dynamic system state matrix |
| $B$ | System input matrix |
| $S$ | Noise input matrix |
| $x$ | System state variables |
| $u$ | System input variables |
| $R(\tau)$ | Stationary $\tau$–lag correlation matrix |
| $C$ | Stationary covariance matrix |

This work was supported by the Natural Sciences and Engineering Research Council (NSERC) under Discovery Grants RGPIN-2016-69152 and RGPIN-2016-04570, and the Fonds de Recherche du Quebec-Nature et technologies under Grant FRQ-NT PR-253686.

J. Guo, X. Wang and B. Ooi are with the Department of Electrical and Computer Engineering, McGill University, 3480 University Street, Montreal, Canada. (email: jinpeng.guo@mail.mcgill.ca; xiaozhe.wang2@mcgill.ca; boon-teck.ooi@mcgill.ca)

## I. INTRODUCTION

With an increasing penetration of renewable energy sources (RES), a growing number of VSCs have been deployed in power grids, serving as the interface between the RES and the AC systems. However, the interactions between VSCs and the main power system dynamics (e.g., the dynamics of synchronous generators) need to be carefully studied to avoid any potential adverse impacts [1] and to improve the system damping performance using the supplementary control (e.g., [2]).

To evaluate the impact of VSCs on the stability of AC power systems, the most straightforward approach is to linearize the dynamical power system model integrating VSCs and to conduct the eigenvalue analysis based on the dynamic system state matrix (e.g., [3]). However, such approach relies heavily on the accuracy of system dynamic model, network topology, VSCs' control strategy, and all the parameter values, which may not be available in real-life applications due to malfunctions of measurement devices, communication failures, etc. [4]. The wide deployment of phasor measurement units (PMUs), on the other hand, provides a new opportunity to develop measurement-based methods to monitor system dynamics, to conduct stability assessments, and to design damping controllers. Measurement-based methods such as the frequency domain decomposition [5], the transfer function identification [6] and the stochastic subspace method [7] have been proposed for the mode identification of conventional power systems. Particularly, for power systems integrating power electronic devices (e.g., HVDC), the authors of [8] have developed a PMU-based method of mode identification based on a system identification method. However, all the aforementioned measurement-based methods may not estimate the true system dynamic state matrix (subject to the similarity transformation) [4], resulting in a great challenge in damping control. To address this issue, our previous work [9] have proposed a hybrid measurement-based and model-based method for estimating the true dynamic system state matrix in ambient condition, which can be used in mode identification [10] and wide-area damping control [11]. A fully model-free method for estimating the dynamic system state matrix is further developed in [12]. Nevertheless, the method considers only the dynamic model of the conventional AC power system without the integration of VSCs.

In this paper, the integration of VSCs requires relatively sophisticated mathematical manipulation to convert the complete model described by differential algebraic equation into a state-space representation, which is not something direct and obvious from [12]. More importantly, to the best knowledge of authors, the proposed method seems to be first model-free method that can estimate the participation factors of oscillation modes for power systems with integrated VSCs. The estimated participation factors will provide significant information to design damping controllers using VSCs, which will be investigated in the near future.

The rest of the paper is organized as follows. Section II integrates the VSC model into the state-space representation of the AC systems. Section III presents the measurement-based method to estimate the dynamic system state matrix and identify the electromechanical modes. Section IV presents the numerical studies in the IEEE 68-bus system to validate the proposed method. Section VI summarizes the conclusions.

## II. SYSTEM MODELING

### A. Stochastic Dynamic AC power systems

In this paper, we follow the approach adopted in [13] to model the loads as constant impedances experiencing Gaussian variations, which can be reflected in the diagonal elements of the reduced admittance matrix $Y(i,i) = G_{ii} + B_{ii} = Y_{ii}(1+\sigma_i\xi_i)\angle\phi_{ii}$, where $\xi_i$ is a standard Gaussian variable and $\sigma_i^2$ describes the strength of the fluctuations.

Dynamics of the generator can be described by [9]

$$\dot{\boldsymbol{\delta}} = \omega_0(\boldsymbol{\omega}-\boldsymbol{1})$$
$$M\dot{\boldsymbol{\omega}} = \boldsymbol{P_M} - \boldsymbol{P_E} - D(\boldsymbol{\omega}-\boldsymbol{1}) - E^2 G\Sigma\boldsymbol{\xi} \quad (1)$$

where $\boldsymbol{\delta} = [\delta_1,\delta_2,...,\delta_{N_g}]^T$ is the vector of generator rotor angles, $\boldsymbol{\omega} = [\omega_1,\omega_2,...,\omega_{N_g}]^T$ is the vector of generator rotor speeds and $\omega_0$ is the base value of speed, $M = diag([M_1,M_2,...,M_{N_g}])$ is the inertia coefficient matrix, $D = diag([D_1,D_2,...,D_{N_g}])$ is the damping coefficient matrix, $\boldsymbol{P_M} = [P_{m1},P_{m2},...,P_{mN_g}]^T$ is the mechanical power input of generators, $\boldsymbol{P_E} = [P_{e1},P_{e2},...,P_{eN_g}]^T$ is the electromagnetic output power of generators, $E = diag([E_1,E_2,...,E_{N_g}])$ is the constant generator voltage behind the transient reactance, $G = diag([G_{11},G_{22},...,G_{N_gN_g}])$ is the equivalent conductance of the system reduced to generator buses, $\Sigma = diag([\sigma_1,\sigma_2,...,\sigma_{N_g}])$, $\boldsymbol{\xi} = [\xi_1,\xi_2,...,\xi_{N_g}]^T$ and $N_g$ is the number of generators.

Linearization of (1) gives

$$\Delta\dot{\boldsymbol{\delta}} = \omega_0\Delta\boldsymbol{\omega}$$
$$M\Delta\dot{\boldsymbol{\omega}} = -\Delta\boldsymbol{P_E} - D\Delta\boldsymbol{\omega} - E^2 G\Sigma\boldsymbol{\xi} \quad (2)$$

### B. Stochastic Dynamic AC Power Systems Integrating VSCs

In order to integrate VSCs into the AC system, all buses other than the generator buses and VSC buses are eliminated to get a simplified system [14], as shown in Fig. 1.

Considering the simplified power system with $N_g$ generators and $N_v$ VSCs, linearization of the power injection equations gives

$$\begin{bmatrix}\Delta\boldsymbol{P_E}\\ \Delta\boldsymbol{P_v}\\ \Delta\boldsymbol{Q_v}\end{bmatrix} = \begin{bmatrix}A_{11} & A_{12} & A_{13}\\ A_{21} & A_{22} & A_{23}\\ A_{31} & A_{32} & A_{33}\end{bmatrix}\begin{bmatrix}\Delta\boldsymbol{\delta}\\ \Delta\boldsymbol{\theta}\\ \Delta\boldsymbol{V}\end{bmatrix} = \begin{bmatrix}\frac{\partial\boldsymbol{P_E}}{\partial\boldsymbol{\delta}} & \frac{\partial\boldsymbol{P_E}}{\partial\boldsymbol{\theta}} & \frac{\partial\boldsymbol{P_E}}{\partial\boldsymbol{V}}\\ \frac{\partial\boldsymbol{P_v}}{\partial\boldsymbol{\delta}} & \frac{\partial\boldsymbol{P_v}}{\partial\boldsymbol{\theta}} & \frac{\partial\boldsymbol{P_v}}{\partial\boldsymbol{V}}\\ \frac{\partial\boldsymbol{Q_v}}{\partial\boldsymbol{\delta}} & \frac{\partial\boldsymbol{Q_v}}{\partial\boldsymbol{\theta}} & \frac{\partial\boldsymbol{Q_v}}{\partial\boldsymbol{V}}\end{bmatrix}\begin{bmatrix}\Delta\boldsymbol{\delta}\\ \Delta\boldsymbol{\theta}\\ \Delta\boldsymbol{V}\end{bmatrix} \quad (3)$$

where $\boldsymbol{P_v} = [P_{v1},P_{v2}\cdots P_{vN_v}]^T$ and $\boldsymbol{Q_v} = [Q_{v1},Q_{v2}\cdots Q_{vN_v}]^T$ are the real-time active and reactive power references of the VSCs. $\boldsymbol{\theta}$ and $\boldsymbol{V}$ are the voltage angles and magnitudes of VSC buses.

According to the second row and the third row of (3), $\Delta\boldsymbol{\theta}$ and $\Delta\boldsymbol{V}$ can be expressed by

$$\Delta\boldsymbol{\theta} = B_1^{-1}\left[A_{23}^{-1}(\Delta\boldsymbol{P_v} - A_{21}\Delta\boldsymbol{\delta}) - A_{33}^{-1}(\Delta\boldsymbol{Q_v} - A_{31}\Delta\boldsymbol{\delta})\right]$$
$$\Delta\boldsymbol{V} = B_2^{-1}\left[A_{22}^{-1}(\Delta\boldsymbol{P_v} - A_{21}\Delta\boldsymbol{\delta}) - A_{32}^{-1}(\Delta\boldsymbol{Q_v} - A_{31}\Delta\boldsymbol{\delta})\right] \quad (4)$$

where $B_1 = A_{23}^{-1}A_{22} - A_{33}^{-1}A_{32}$ and $B_2 = A_{22}^{-1}A_{23} - A_{32}^{-1}A_{33}$.

Substituting $\Delta\boldsymbol{\theta}$ and $\Delta\boldsymbol{V}$ from (4) to the first row of (3), the generators output power can be expressed a function of

$$\Delta\boldsymbol{P_E} = A_{11}\Delta\boldsymbol{\delta} + A_{12}\Delta\boldsymbol{\theta} + A_{13}\Delta\boldsymbol{V}$$
$$= A_1\Delta\boldsymbol{\delta} + A_2\Delta\boldsymbol{P_v} + A_3\Delta\boldsymbol{Q_v} \quad (5)$$

where

$$A_1 = A_{11} + A_{12}B_1^{-1}\left(-A_{23}^{-1}A_{21} + A_{33}^{-1}A_{31}\right) + A_{13}B_2^{-1}\left(-A_{22}^{-1}A_{21} + A_{32}^{-1}A_{31}\right),$$
$$A_2 = A_{12}B_1^{-1}A_{23}^{-1} + A_{13}B_2^{-1}A_{22}^{-1},$$
$$A_3 = -A_{12}B_1^{-1}A_{33}^{-1} - A_{13}B_2^{-1}A_{32}^{-1}.$$

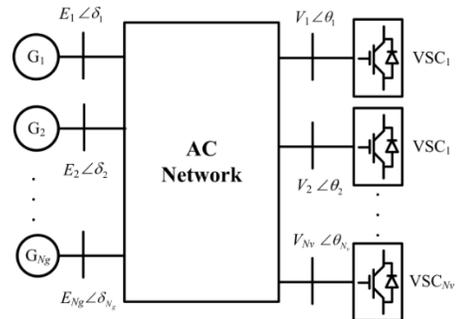

Fig. 1. AC power system integrated with VSCs

Substitution of $\Delta P_E$ from (5) into (2) gives

$$\dot{x} = Ax + Bu + S\xi \tag{6}$$

where $x = [\Delta\delta, \Delta\omega]^T$, $u = [\Delta P_v, \Delta Q_v]^T$, $S = [0, -M^{-1}E^2G\Sigma]^T$,

$A = \begin{bmatrix} 0 & \omega_0 I_{N_g} \\ -M^{-1}A_1 & -M^{-1}D \end{bmatrix}$ and $B = \begin{bmatrix} 0 & 0 \\ -M^{-1}A_2 & -M^{-1}A_3 \end{bmatrix}$.

*C. VSC model*

Since electromechanical oscillations are considered, the VSC converters are simply regarded as power sinks and sources similar to [2]. Power oscillation damping support from VSCs can be provided by adding supplementary active and reactive power signals to the steady-state references [15]. Hence, the power references of VSCs are controlled by

$$\begin{aligned} P_v &= P_{vs} + P_{vd} \\ Q_v &= Q_{vs} + Q_{vd} \end{aligned} \tag{7}$$

where $P_{vs}$ and $Q_{vs}$ are the steady-state active and reactive power references; $P_{vd}$ and $Q_{vd}$ are the supplementary damping power references. If the frequency variation is used as the feedback signals [2], we have

$$\begin{aligned} P_{vd} &= K_1(\omega - 1) \\ Q_{vd} &= K_2(\omega - 1) \end{aligned} \tag{8}$$

where $K_1$ and $K_2$ are damping coefficients of the active and reactive power control of VSCs.

Linearization of (7) and (8) gives

$$\begin{aligned} \Delta P_v &= \Delta P_{vd} = K_1 \Delta\omega \\ \Delta Q_v &= \Delta Q_{vd} = K_2 \Delta\omega \end{aligned} \tag{9}$$

Substituting $\Delta P_v$ and $\Delta Q_v$ in (6) by (9) gives

$$\dot{x} = A_c x + S\xi \tag{10}$$

where $A_c = \begin{bmatrix} 0 & \omega_0 I_{N_g} \\ -M^{-1}A_1 & -M^{-1}(A_2 K_1 + A_3 K_2 + D) \end{bmatrix}$.

The closed-loop dynamic system state matrix $A_c$ carries significant information about the system's dynamic states, which can be used in the stability analysis such as identifying electromechanical modes. We will present a PMU-based method for estimating $A_c$ in the following section.

## III. MEASUREMENT-BASED ESTIMATON OF SYSTEM STATE MATRIX

Specifically, the vector of state variables $x$ described in (10) is a multivariate Ornstein-Uhlenbeck (OU) process. According to the regression theorem of OU process [16], if $A_c$ is stable that is typically satisfied in ambient conditions, the following relation holds:

$$\frac{d}{d\tau}[R(\tau)] = A_c R(\tau) \tag{11}$$

where $R(\tau)$ is the $\tau$-lag correlation matrix and is defined by

$$R(\tau) = \left\langle [x(t+\tau) - \bar{x}][x(t) - \bar{x}]^T \right\rangle.$$

Therefore, the dynamic system state matrix can be estimated from [12]:

$$A_c = \frac{1}{\tau}\log[R(\tau)C^{-1}] \tag{12}$$

where $C$ is the stationary covariance matrix and is defined as

$$C = \left\langle [x(t) - \bar{x}][x(t) - \bar{x}]^T \right\rangle.$$

The correlation matrix in (11) and the stationary covariance matrix in (12) are typically unknown in practice, but they can be estimated from sufficient PMU data. Each entry of the sample $\tau$-lag correlation matrix and the sample covariance matrix can be calculated as follows:

$$\begin{aligned} \hat{R}_{\delta_i \delta_j}(\tau) &= \frac{1}{N}\sum_{k=1}^{N-\tau}(\delta_{(k+\tau)i} - \bar{\delta_i})(\delta_{kj} - \bar{\delta_j}) \\ \hat{C}_{\delta_i \delta_j} &= \frac{1}{N}\sum_{k=1}^{N}(\delta_{ki} - \bar{\delta_i})(\delta_{kj} - \bar{\delta_j}) \end{aligned} \tag{13}$$

where $\tau$ is the lagging time steps, $\bar{\delta_i}$ is the mean value of $\delta_i$, and $N$ is the sample size.

As a result, the system state matrix can be estimated by

$$A_c = \frac{1}{\Delta t}\log\left[\hat{R}(\tau)\hat{C}^{-1}\right] \tag{14}$$

where $\Delta t$ is the sampling time.

Equation (14) subtly connects the physical model knowledge with the statistic properties of PMU data, serving as the theoretic basis of the proposed measurement-based method for electromechanical mode estimation for AC power systems with integrated VSCs.

We assume that all generator buses are equipped with PMUs. Such assumption may be too optimistic in current power systems, yet is foreseeable in the near future given the increasing investment of PMU technologies worldwide. In addition, we assume that the rotor angle $\delta$ and the rotor angular velocity $\omega$ can be estimated from PMU data in ambient conditions as shown in [17]. As such, the following algorithm can estimate the dynamic system state matrix and identify electromechanical modes for AC power systems with VSCs using only PMU data.

**Step 1:** Given PMU data with a sufficient window length, estimate $\delta$ and $\omega$ from PMU measurements and calculate the sample correlation matrix $\hat{R}(\tau)$ and the sample stationary covariance $\hat{C}$ by (13).

**Step 2:** Estimate the dynamic system state matrix $A_c$ by (14).

**Step 3:** Identify the electromechanical modes (e.g., estimate oscillation frequencies, mode shapes and participation factors) for the AC power system with integrated VSCs by calculating $A_c$'s eigenvalues, the eigenvectors, etc.

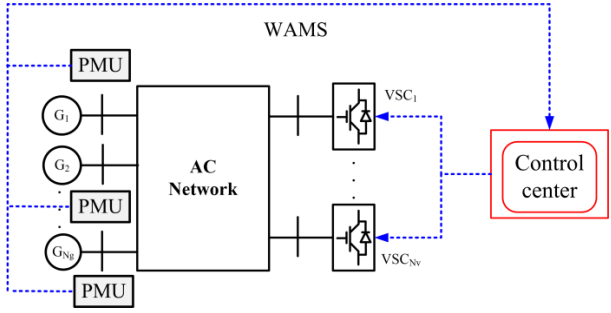

Fig. 2. Control scheme of the online estimation strategy

The system-level control mechanism is illustrated in Fig. 2. The data is collected at the generator side and the system dynamics are monitored by the control center, based on which VSCs can be controlled. Note that the proposed method is purely data-driven, requiring no knowledge about the dynamic model and the network model. The emulated PMU data with a window length of 300s is used in this paper, which presents a good accuracy in the estimation results. More discussions on the impact of window length on estimation accuracy could be found in [12].

## IV. CASE STUDIES

We consider the IEEE 68-bus system, in which three VSC stations are added to bus 54, 17 and 42, respectively. The reference power $P_{vs}=0.5$, $Q_{vs}=0$ are set for all VSCs. To validate the proposed PMU-based method for estimating the system state matrix of AC power systems with VSCs, two cases are considered. In the first case, VSCs work only as constant power sources, without providing supplementary control ($K_1=0$, $K_2=0$ in (8)). In the second case, the damping coefficients of VSCs are adjusted to provide damping. It will be shown that accurate estimation results can be achieved in both cases, showing the correctness of the formulation of the estimation when integrating VSCs and the accuracy of the proposed method. Also, improved damping performance can be achieved by adjusting the damping coefficients $K_1$ and $K_2$, which sheds light on designing damping controllers using VSCs in the future.

Although the proposed method is developed using the classical generator model, the 3rd-order generator models are used in all simulations to show the feasibility of the proposed method in practical applications. Moreover, G1-G12 are controlled by automatic voltage regulators (AVRs). The stochastic fluctuation intensities $\sigma_1,...,\sigma_n$ in (1) are set to be 0.05. A sampling rate of 50Hz and a window size of 300s are used to generate the emulated PMU data. Sample stochastic trajectories are given in Fig. 4. All the simulations are conducted using the Power System Analysis Toolbox (PSAT) [18].

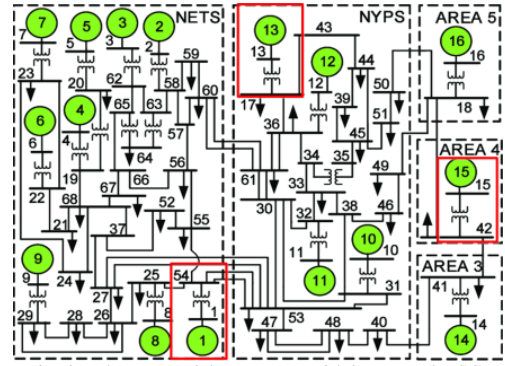

Fig. 3. The IEEE 68-bus system with integrated VSCs

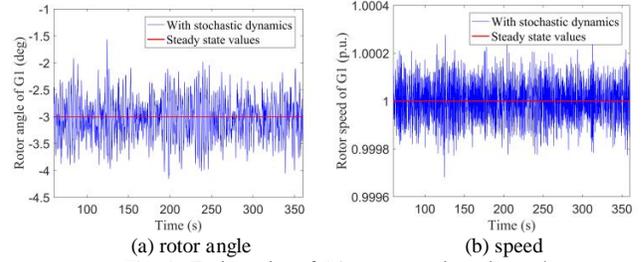

(a) rotor angle (b) speed

Fig. 4. Trajectories of G1's rotor angle and speed

### A. Case I: VSCs work as constant power sources

In the first case, the supplementary control of VSCs is not implemented, i.e., $K_1=0$, $K_2=0$ in (8). We applied the proposed measurement-based method for estimating the dynamic system state matrix and the electromechanical modes. The estimated frequencies $f_e$ and damping ratios $\xi_e$ of all oscillation modes are compared with their actual values $f_a$ and $\xi_a$, as presented in Table I. It can be seen that the estimation errors for most modes are below 10%, showing that the proposed method can accurately identify the modes.

Table I. A comparison between the actual and estimated values of frequencies and damping ratios of different oscillation modes (Case I)

| Mode | Frequency | | | Damping ratio | | |
|---|---|---|---|---|---|---|
| | $f_a$ (Hz) | $f_e$ (Hz) | Error (%) | $\xi_a$ (%) | $\xi_e$ (%) | Error (%) |
| 1 | 0.384 | 0.389 | 1.27 | 9.055 | 8.826 | -2.53 |
| 2 | 0.595 | 0.597 | 0.46 | 7.983 | 7.187 | -9.97 |
| 3 | 0.957 | 0.960 | 0.33 | 5.092 | 4.799 | -5.75 |
| 4 | 1.107 | 1.105 | -0.14 | 4.792 | 4.456 | -7.02 |
| 5 | 1.717 | 1.711 | -0.31 | 5.998 | 6.168 | 2.84 |
| 6 | 1.781 | 1.760 | -1.17 | 3.054 | 2.857 | -6.45 |
| 7 | 1.968 | 1.948 | -1.01 | 3.198 | 3.119 | -2.47 |
| 8 | 2.084 | 2.071 | -0.61 | 1.646 | 1.566 | -4.84 |
| 9 | 2.205 | 2.190 | -0.69 | 1.361 | 1.464 | 7.62 |
| 10 | 2.250 | 2.243 | -0.32 | 1.616 | 1.507 | -6.75 |
| 11 | 2.371 | 2.364 | -0.31 | 0.993 | 0.938 | -5.50 |
| 12 | 2.649 | 2.620 | -1.12 | 2.511 | 2.478 | -1.34 |
| 13 | 2.652 | 2.635 | -0.62 | 1.874 | 1.744 | -6.96 |
| 14 | 2.779 | 2.766 | -0.45 | 1.059 | 1.042 | -1.56 |
| 15 | 2.816 | 2.806 | -0.35 | 1.810 | 2.065 | 14.08 |

Furthermore, the mode shapes can be obtained from the right eigenvectors of the estimated matrix. In particular, the actual and the estimated mode shapes of mode 3 (an inter-area mode highlighted in Table I) are plotted in Fig. 5, showing that the generators in NETS are oscillating against those in NYPS. The estimated participation factors of all generators in mode 3 are also compared with the true values in Fig. 6. It can be observed that both the mode shape and the participation factor can be accurately estimated. Since the participation factors carry crucial information in the control design, effective damping controllers may be further developed to suppress critical interarea modes [11].

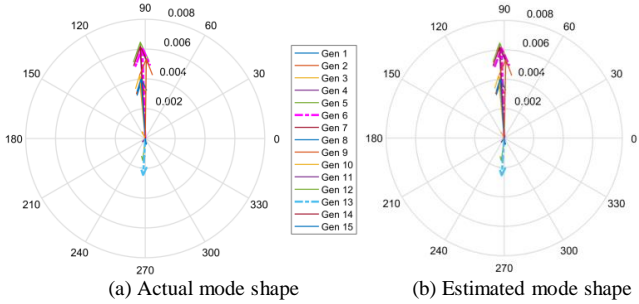

(a) Actual mode shape    (b) Estimated mode shape
Fig. 5. The actual and estimated mode shapes of oscillation mode 3

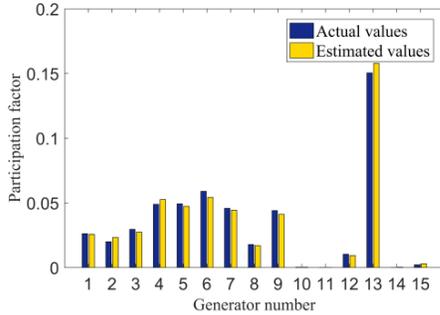

Fig. 6. The actual and estimated participation factors of oscillation mode 3

### B. Case II: VSCs provide supplementary damping control

In order to check the impact of VSCs' control strategy on system stability, the damping coefficient $K_1$ in (8) is adjusted in this case. According to the analysis of participation factor as shown in Fig. 6, both G6 and G13 play significant roles in the oscillation mode 3. While G13 is electrically close to VSC2 (at Bus 17), the control of VSC2 may be effective in suppressing the oscillation mode 3. We, therefore, set entries $K_1(2,6)$ and $K_1(2,13)$ of the damping coefficient $K_1$ to be 300 and -300, respectively, indicating that the frequency deviations of G6 and G13 are used as feedbacks for VSC. Note that the opposite sign is selected because G6 and G13 are oscillating against each other as can be seen in the mode shape shown in Fig. 5 (dashed lines). The adjustment of $K_1$ changes the lower right submatrix of $A_c$, as highlighted in Fig. 7 that compares the estimated values of the lower right submatrix of $A_c$ with and without the damping control.

The comparison between the actual and the estimated frequencies as well as the damping ratios of the new oscillation modes after implementing the control are presented in Table II. Again, a good estimation accuracy is achieved. Compared to the results in Table I, the actual damping ratio of oscillation mode 3 (also marked in yellow in Table II) is increased from 5.092% to 9.096%, while the proposed method can accurately capture the increase of the damping ratio because of the supplementary control of VSCs.

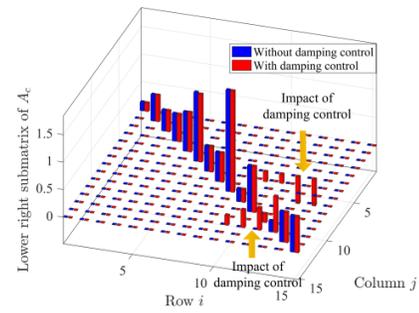

Fig. 7. The estimated values of the lower right submatrix of $A_c$ with and without damping control

Table II. A comparison between the actual and estimated values of frequencies and damping ratios of different oscillation modes (Case II)

| Mode | Frequency | | | Damping ratio | | |
|---|---|---|---|---|---|---|
| | $f_a$ (Hz) | $f_e$ (Hz) | Error (%) | $\xi_a$ (%) | $\xi_e$ (%) | Error (%) |
| 1 | 0.383 | 0.389 | 1.40 | 8.484 | 7.878 | -7.14 |
| 2 | 0.594 | 0.597 | 0.43 | 8.052 | 7.184 | -10.79 |
| 3 | 0.958 | 0.960 | 0.28 | 9.096 | 9.298 | 2.22 |
| 4 | 1.107 | 1.106 | -0.12 | 4.792 | 4.448 | -7.17 |
| 5 | 1.717 | 1.713 | -0.25 | 5.944 | 6.074 | 2.19 |
| 6 | 1.780 | 1.759 | -1.19 | 3.030 | 2.762 | -8.84 |
| 7 | 1.968 | 1.949 | -1.00 | 3.197 | 3.070 | -3.98 |
| 8 | 2.084 | 2.071 | -0.61 | 1.639 | 1.562 | -4.70 |
| 9 | 2.204 | 2.189 | -0.68 | 1.311 | 1.403 | 7.02 |
| 10 | 2.250 | 2.243 | -0.32 | 1.617 | 1.505 | -6.89 |
| 11 | 2.371 | 2.364 | -0.30 | 0.993 | 0.937 | -5.63 |
| 12 | 2.649 | 2.620 | -1.12 | 2.511 | 2.449 | -2.47 |
| 13 | 2.652 | 2.635 | -0.63 | 1.874 | 1.741 | -7.08 |
| 14 | 2.779 | 2.766 | -0.45 | 1.059 | 1.037 | -2.13 |
| 15 | 2.815 | 2.806 | -0.34 | 1.809 | 2.063 | 14.07 |

It should also be noted that the controller design, i.e., $K_1$ and $K_2$, is not optimized, yet is tuned intuitively from the participation factor to demonstrate the potential of using VSCs in damping improvement, which is also among our future work.

### C. Impact of measurement noises

Table III. A comparison between the actual and estimated values of frequencies and damping ratios with measurement noises (Case II)

| Mode | Frequency | | | Damping ratio | | |
|---|---|---|---|---|---|---|
| | $f_a$ (Hz) | $f_e$ (Hz) | Error (%) | $\xi_a$ (%) | $\xi_e$ (%) | Error (%) |
| 1 | 0.383 | 0.389 | 1.36 | 8.484 | 7.656 | -9.75 |
| 2 | 0.594 | 0.597 | 0.43 | 8.052 | 7.277 | -9.62 |
| 3 | 0.958 | 0.961 | 0.31 | 9.096 | 9.298 | 2.22 |
| 4 | 1.107 | 1.105 | -0.15 | 4.792 | 4.600 | -4.00 |
| 5 | 1.717 | 1.713 | -0.25 | 5.944 | 6.075 | 2.21 |
| 6 | 1.780 | 1.759 | -1.20 | 3.030 | 2.782 | -8.18 |
| 7 | 1.968 | 1.948 | -1.01 | 3.197 | 3.070 | -3.97 |
| 8 | 2.084 | 2.071 | -0.61 | 1.639 | 1.588 | -3.13 |
| 9 | 2.204 | 2.189 | -0.68 | 1.311 | 1.397 | 6.58 |
| 10 | 2.250 | 2.243 | -0.32 | 1.617 | 1.510 | -6.58 |
| 11 | 2.371 | 2.364 | -0.30 | 0.993 | 0.947 | -4.65 |
| 12 | 2.649 | 2.620 | -1.12 | 2.511 | 2.447 | -2.54 |
| 13 | 2.652 | 2.635 | -0.63 | 1.874 | 1.732 | -7.60 |
| 14 | 2.779 | 2.766 | -0.45 | 1.059 | 1.045 | -1.32 |
| 15 | 2.815 | 2.806 | -0.35 | 1.809 | 2.185 | 20.83 |

Measurement noises are inevitable in practical applications so that the performance of the method under measurement noise needs to be tested. Particularly, a

Gaussian measurement noise with a standard deviation of $10^{-3}$ was added to the rotor angles and a Gaussian measurement noise with a standard deviation of $10^{-6}$ was added to the rotor speeds [19]. As shown in Table III, the estimation results for the oscillation frequencies and the damping ratios still maintain good accuracy, demonstrating the robustness of the proposed method against measurement noise.

## V. Conclusion and Perspectives

In this paper, we have proposed a measurement-based method to estimate the dynamic system state matrix and identify the electromechanical modes for AC system with integrated VSCs. The case studies in the IEEE 68-bus system show that the proposed method can accurately identify the electromechanical oscillation even with measurement noise. In addition, the tuning of the damping coefficient of VSCs based on the estimated participation factor may improve the system damping performance. The work may shed light on designing measurement-based method to improve the damping performance using VSCs, which is among our future work.